\documentclass{ws-procs9x6}

\begin{document}

\title{BL Lac evolution revisited}

\author{ANNA WOLTER, FRANCESCA CAVALLOTTI}

\address{INAF- Osservatorio Astronomico di Brera\\
Via Brera, 28 20121 Milano, ITALY \\
E-mail: anna@brera.mi.astro.it; cavallot@brera.mi.astro.it}

\author{JOHN T. STOCKE}

\address{Center for Astrophysics and Space Astronomy\\
University of Colorado, Boulder, CO 80309-0389, USA}

\author{TRAVIS RECTOR}

\address{University of Alaska\\
3211 Providence Dr., BMB 212, Anchorage, AK 99508, USA}


\maketitle

\abstracts{BL Lac objects are an elusive and rare class of active galactic
nuclei. For years their evolutionary behavior has appeared
inconsistent with the trend observed in the population of AGN at
large. The so-called ``negative'' evolution implies that BL Lacs were
either less or fainter in the past. This effect is stronger for BL
Lacs selected in X-ray surveys. We have investigated if one of the
selection criteria, namely the flat-radio spectrum (imposed on the
Radio-selected but not on the X-ray-selected samples), might explain
the different evolutionary trend.
}

\section{BL Lacs samples and evolution}

Radio selected and X-ray selected samples of BL Lac objects are
created through various selections and show
different evolutionary properties as shown in Table~\ref{table1}. 
It is evident from the Table indications that only samples that impose
the $\alpha_r$ (flat radio spectrum) criterion show positive 
evolution (vs. negative or no-evolution).
We want to analyze whether this selection criterion is a primary 
cause for the different evolution measured in different sample.
We obtained simultaneous data at the Very Large Array
(VLA\footnote{The VLA is a facility of the National Radio Astronomy
Observatory.  The National Radio Astronomy Observatory (NRAO) is a
facility of the National Science Foundation operated under
cooperative agreement by Associated Universities, Inc.})  at 20, 6 and
3.6 cm (C conf.) for a
(radio) flux limited sample of 26 sources in the EMSS
([\cite{em1}],[\cite{em2}]) with the purpose of studying the
distribution of $\alpha_r$ in a complete unbiased sample.  A detailed
analysis is available in [\cite{cwsr}].

\begin{table}[htb]
\tbl{Summary of BL Lacs samples (their size), measure of evolution and 
selection criteria applied}
{\footnotesize
\begin{tabular}{@{}lrr@{}}
\hline
\hline
Samples (\#) &         Evolution is:  &	Selection criteria applied \\[1ex]
\hline
{\em X-ray selected}  & & \\[1ex]
EMSS  (40 objects)    & negative   & 		f$_X$  \\[1ex]
{\em Radio selected}  & & \\[1ex]
1Jy (34 objects)      &	{\em positive}/null &	f$_r$ - mag - {\bf $\alpha_r$} \\[1ex]
{\em Radio \& X-ray selected}  & & \\
RGB (33 objects)      & negative  &	f$_X$ - f$_r$ - mag \\[1ex]
DXRBS (30 objects)    &	{\em positive}  &	f$_X$ - f$_r$ - {\bf $\alpha_r$} \\ [1ex]
REX (55 in XB-REX)    & negative/null &	f$_X$ - f$_r$ - mag \\[1ex]
HRX (77 objects)      & negative  & 	f$_X$ - f$_r$ \\[1ex]
{\em Others}  	      & & \\[1ex]
Sedentary (58 objs)   &	negative  & 	f$_X$ - f$_r$ - mag -$\alpha_{ro}$ $\alpha_{rx}$ - (HR) \\[1ex]
[PG - (6 objs)]	      & unknown	  & 	UV excess \\[1ex]
\hline
\hline
\end{tabular}\label{table1} }
\begin{tabnote}
EMSS: Rector et al. 2000 AJ 120 1626; 1Jy: Stickel et al. 1991 ApJ 374 431; 
RGB: Laurent-Muehleisen et al. 1993 AJ 106 875; DXRBS: Landt et al. 2001 
MNRAS 323 757; REX: Caccianiga et al. 2002 ApJ 566 181;
HRX: Beckmann et al. 2003 A\&A 411 327; Sedentary: Giommi et al. 1999
MNRAS 310 65; PG: Schmidt \& Green 1983, ApJ 269 352.
\end{tabnote}
\vspace*{-5pt}
\end{table}

\smallskip
\noindent
We find that a large fraction of BL Lacs show a steep spectrum 
(15-40\% depending on conditions of observations). 
These objects are missed in radio surveys that need an $\alpha_r$ 
selection criterion to avoid the bulk of radio galaxies.
However, on average, steep spectrum sources have similar properties 
to flat spectrum sources.  We conclude that 
the spectral selection is not responsible for the different evolution 
properties of RBL and XBL.

\section*{Acknowledgments}
This work has received partial financial support from the Italian
Space Agency (ASI) and MIUR.

%
%
%
%

\end{document}